# High pressure synthesis of FeO–ZnO solid solutions with rock salt structure: *in situ* X-ray diffraction studies

Petr S. Sokolov[1,2], Andrey N. Baranov[3], Christian Lathe[4], Vladimir L. Solozhenko[1*]

[1] *LPMTM-CNRS, Université Paris Nord, 93430 Villetaneuse, France*

[2] *Materials Science Department, Moscow State University, 119991 Moscow, Russia*

[3] *Chemistry Department, Moscow State University, 119991 Moscow, Russia*

[4] *HASYLAB-DESY, 22603 Hamburg, Germany*

**Abstract**

X-ray diffraction with synchrotron radiation has been used for the first time to study chemical interaction in the FeO–ZnO system at 4.8 GPa and temperatures up to 1300 K. Above 750 K, the chemical reaction between FeO and ZnO has been observed that resulted in the formation of rock salt (*rs*) $Fe_{1-x}Zn_xO$ solid solutions ($0.3 \leq x \leq 0.85$). The lattice parameters of these solid solutions have been *in situ* measured as a function of temperature under pressure, and corresponding thermal expansion coefficients have been calculated.

* e-mail: vladimir.solozhenko@univ-paris13.fr

**Introduction**

Zinc oxide is a wide-band-gap semiconductor and has many industrial applications. Recently ZnO-based solid solutions with transition metal oxides have gained substantial interest as promising magnetic semiconductors [1]. At ambient pressure ZnO has hexagonal wurtzite structure (P6$_3$mc) that transforms into rock salt one (Fm3m) at pressures above 6 GPa [2,3]. However, high-pressure phase of ZnO cannot be quenched down to ambient conditions [3]. Very recently, metastable Me$^{II}$O-ZnO solid solutions (Me$^{II}$ – Ni$^{2+}$, Fe$^{2+}$, Co$^{2+}$, Mn$^{2+}$) with rock salt structure have been synthesized by quenching from 7.7 GPa and 1450-1650 K [4]; in particular, single-phase rock salt Fe$_{1-x}$Zn$_x$O solid solutions of various stoichiometries ($0 < x \leq 0.5$) have been obtained. At ambient pressure equilibrium phase diagram of the FeO-ZnO system is of peritectic type with a wide (from 20 to 80 wt% ZnO at 1073 K) region of coexistence of rock salt FeO-based and wurtzite ZnO-based solid solutions [5]. However, no information on chemical interaction and phase relations in the system at high pressures and temperatures can be found in the literature. In the present work we have performed the first *in-situ* investigation of the binary FeO-ZnO system at high pressures and temperatures using X-ray diffraction with synchrotron radiation.

**Experimental**

Powders of wurtzite ZnO (Alfa Aesar, 99.99%, 325 mesh) and rock salt "FeO" (Aldrich, 99%, 10 mesh) have been used as starting materials. The FeO-ZnO mixtures of various stoichiometries (30, 50, 70 and 85 mol% ZnO) were thoroughly ground in a mortar, compressed into disks and placed into capsules of high-purity hexagonal graphite-like boron nitride (hBN).

High-pressure experiments have been performed using MAX80 multianvil X-ray system at beamline F2.1, DORIS III (HASYLAB-DESY). The experimental details and high-pressure setup have been described elsewhere [6,7]. Energy-dispersive X-ray diffraction patterns were collected on a Canberra solid state Ge-detector with fixed Bragg angle $2\theta = 9.073(1)°$ using a white beam collimated down to $100 \times 100$ µm$^2$. The detector was calibrated using the K$_\alpha$ and K$_\beta$ fluorescence lines of Rb, Mo, Ag, Ba, and Tb.

The sample temperature was measured by a Pt10%Rh-Pt thermocouple. The correction for the pressure effect on the thermocouple emf was made using the data of Getting and Kennedy [8] extrapolated to 5 GPa. The temperature of the high-pressure cell was controlled by a Eurotherm PID regulator within ±3 K. Pressures at different temperatures were evaluated from the lattice parameters of highly ordered ($P_3 = 0.98\pm0.02$)[1] hBN using its thermoelastic equation of state [6]. With the increase in temperature from ambient to 1400 K, the pressure increase in the central part of the cell did not exceed 0.2 GPa.

The samples were compressed to the required pressure at ambient temperature, and then diffraction patterns were collected in the "autosequence" mode at a linear heating with a rate of

---

[1] ($P_3 = 1-\gamma$), where $\gamma$ is the concentration of turbostratic stacking faults [9]

10 K/min. With the storage ring operating at 4.44 GeV and 120±30 mA, the time of data collection for each pattern was 60 s.

**Results and Discussion**

At 4.8 GPa and temperatures below 750 K only reflections of pristine oxides (*w*-ZnO and *rs*-FeO) are observed in the diffraction patterns. At higher temperatures intensities of *w*-ZnO reflections start to decrease, while intensities of the cubic phase reflections start to increase, that is indicative of the chemical interaction between FeO and ZnO. This reaction is accompanied by dissolution of zinc oxide in iron (II) oxide and formation of a $Fe_{1-x}Zn_xO$ solid solution with rock salt structure. Regardless of the stoichiometry, the lattice parameters of forming solid solutions are very close to the lattice parameter of pristine FeO,[2] so no appreciable shift of cubic phase reflections is observed in the diffraction patterns, in contrast to the case of the solid solutions formation in the MgO-ZnO system [7].

The onset temperature of interaction, 750 K at 4.8 GPa, was found to be irrespective of the reaction mixture stoichiometry. The forming *rs*-$Fe_{1-x}Zn_xO$ solid solutions coexist with wurtzite ZnO in a rather wide temperature range. Upon temperature increase intensities of *w*-ZnO reflections further decrease (Fig. 1) due to dissolution of zinc oxide in the *rs*-$Fe_{1-x}Zn_xO$ lattice, and finally after complete ZnO dissolution, the solid solution stoichiometry attains the stoichiometry of the initial reaction mixture, and only reflections of the rock salt phase remain in the diffraction patterns (Fig. 1). The temperature of the complete disappearance of *w*-ZnO ($T_d$) depends on the initial mixture stoichiometry as shown in Fig. 1 (inset) and increases with ZnO content from 950(5) K for $x = 0.3$ to 1290(5) K for $x = 0.85$.

At temperatures above 1200 K, drastic changes in mutual intensities of reflections in diffraction patterns are observed for all synthesized rock salt FeO-ZnO solid solutions.[3] Such fluctuations of reflection intensities can be caused by the inferred motion of crystallites due to appearance of a liquid in the system that is typical for an onset of melting. However, according to the phase diagram of the system at ambient pressure, solid solution of limited ZnO solubility in *rs*-FeO at peritectic temperature, $Fe_{0.73}Zn_{0.27}O$, melts at 1734 K [5]. Assuming the same slop of the melting curve as in the case of pristine FeO, 30 K/GPa,[4] one can estimate melting point of this solid solution at 5 GPa as 1900 K, i.e. ~600 K higher than the observed intensity fluctuations in the diffraction patterns of rock salt FeO-ZnO solid solutions. Obviously, this question demands special attention in the future studies of the FeO-ZnO system over the extended range of pressures and temperatures.

Lattice parameters of *rs*-$Fe_{1-x}Zn_xO$ solid solutions ($x = 0.3, 0.5, 0.7$) have been *in situ* determined in the $T_d - 1300$ K temperature range at 4.8 GPa. The temperature dependencies of the unit cell volumes are presented in Fig. 2. All these dependencies are linear, and their slopes give the values of the volume thermal expansion coefficients ($\alpha$) at 4.8 GPa (Tabl. 1). The $\alpha$-value of iron (II) oxide determined in the present work, $5.1(2)\times10^{-5}$ K$^{-1}$, is in a good agreement with the value $4.8(1)\times10^{-5}$ K$^{-1}$ reported in [11]. Linear extrapolation of the thermal expansion coefficients of $rs$-Fe$_{1-x}$Zn$_x$O solid solutions to $x = 1$ (pure ZnO) gives $\alpha$-value of $6.0(6)\times10^{-5}$ K$^{-1}$ which is close to the volume thermal expansion coefficient of $rs$- ZnO, $5.2(2)\times10^{-5}$ K$^{-1}$,[5] reported in [3].

---

[2] At 4.8 GPa and 1200 K the lattice parameters of FeO and $Fe_{0.5}Zn_{0.5}O$ solid solution are 4.315(6) and 4.307(2) Å, respectively.

[3] For *rs*-$Fe_{0.7}Zn_{0.3}O$ an abrupt (more than by one order of magnitude) increase in intensity of *220* reflection is observed, while *331* and *400* reflections disappear almost completely; for $Fe_{0.3}Zn_{0.7}O$, *400* and *420* reflection intensities increase twice, and intensity of *331* reflection decreases. No change in reflection intensities is observed for pristine iron (II) oxide at the same *p,T*-conditions.

[4] Melting point of iron (II) oxide is 1650 K at ambient pressure and ~1800 K at 5 GPa [10].

**Conclusions**

At pressures of about 5 GPa and temperatures above 750 K chemical interaction between FeO and ZnO results in the formation of Fe$_{1-x}$Zn$_x$O ($0.3 \leq x \leq 0.85$) solid solutions with rock salt structure. Formation temperature of single-phase $rs$-Fe$_{1-x}$Zn$_x$O increases with increasing ZnO content from 950 K for $x = 0.3$ to 1290 K for $x = 0.85$. Thermal expansion coefficients calculated from the temperature dependencies of lattice parameters of $rs$-Fe$_{1-x}$Zn$_x$O solid solutions under pressure are intermediate between corresponding values of pristine FeO and ZnO and increase with increasing of the ZnO content.

**Acknowledgements**

We thank Dr. V.Z. Turkevich for his stimulating discussions. Experiments have been performed during beamtime allocated to the Project I-20070033 EC at HASYLAB-DESY and have received funding from the European Community's Seventh Framework Programme (FP7/2007-2013) under grant agreement n° 226716. This work was also supported by the Russian Foundation for Basic Research (Project No 09-03-90442-Укр_ф_а). PSS is grateful to the French Ministry of Foreign Affairs for financial support (BGF fellowship No 2007 1572).

---

[5] Average value of the 300-1273 K temperature and 3.2-10.4 GPa pressure domain.

**Table 1.**

Volume thermal expansion coefficients of the rock salt FeO-ZnO solid solutions at 4.8 GPa

| Composition | "FeO" | $Fe_{0.7}Zn_{0.3}O$ | $Fe_{0.5}Zn_{0.5}O$ | $Fe_{0.3}Zn_{0.7}O$ | $rs$-ZnO* |
|---|---|---|---|---|---|
| $\alpha \times 10^5$ (K$^{-1}$) | 5.1±0.2 | 5.4±0.2 | 5.1±0.4 | 5.9±0.5 | 6.0±0.6 |
| **Temperature range (K)** | 800-1300 | 950-1300 | 1030-1300 | 1170-1300 | – |

* Linear extrapolation of the thermal expansion coefficients of $rs$-$Fe_{1-x}Zn_xO$ solid solutions ($x$ = 0.3, 0.5, 0.7) to $x$ = 1 (pure ZnO)

**Figure captions**

Fig. 1. Diffraction patterns of the FeO-ZnO mixture (50 mol% ZnO) taken at 4.8 GPa in the course of a linear heating at a rate of 10 K/min. Inset: Temperatures of the complete disappearance of wurzite ZnO (solid circles).

Fig. 2. Unit cell volumes of rock salt $Fe_{1-x}Zn_xO$ solid solutions *vs* temperature at 4.8 GPa (squares – $x = 0$, circles – $x = 0.30$, triangles – $x = 0.5$, hexagons – $x = 0.7$).

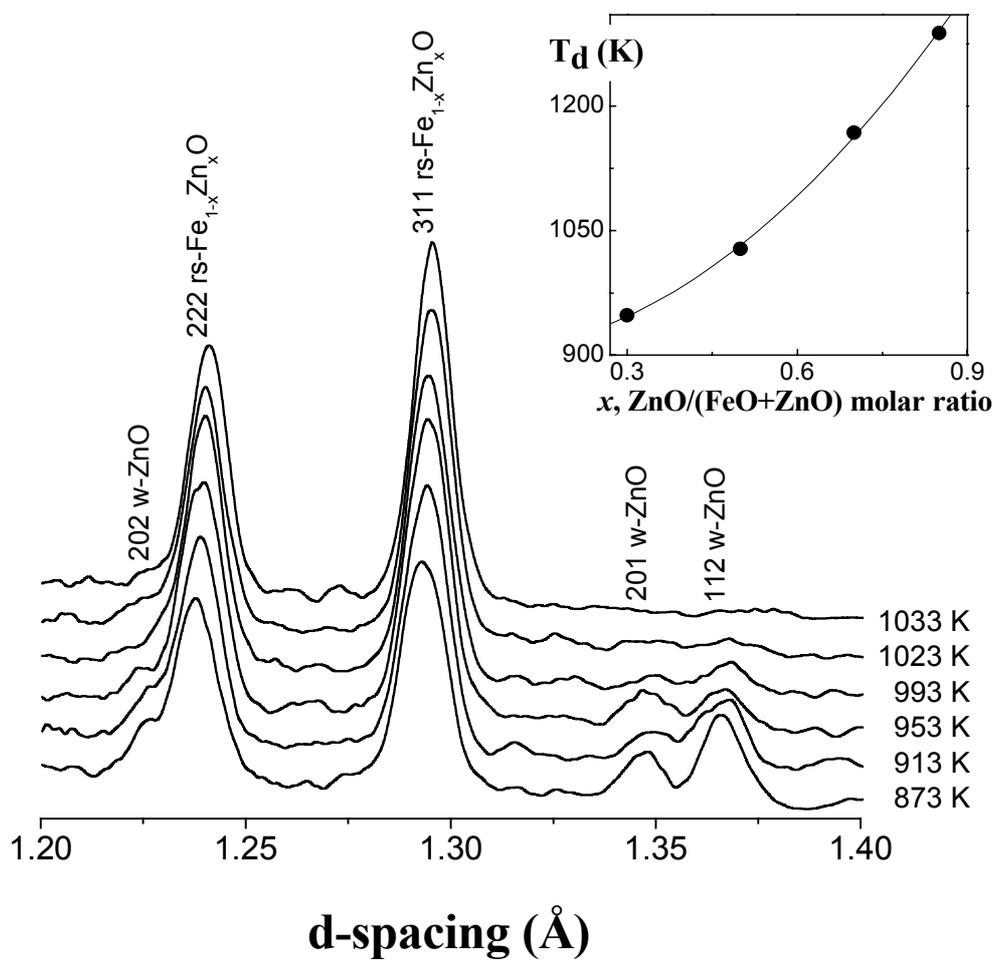

**Fig. 1**

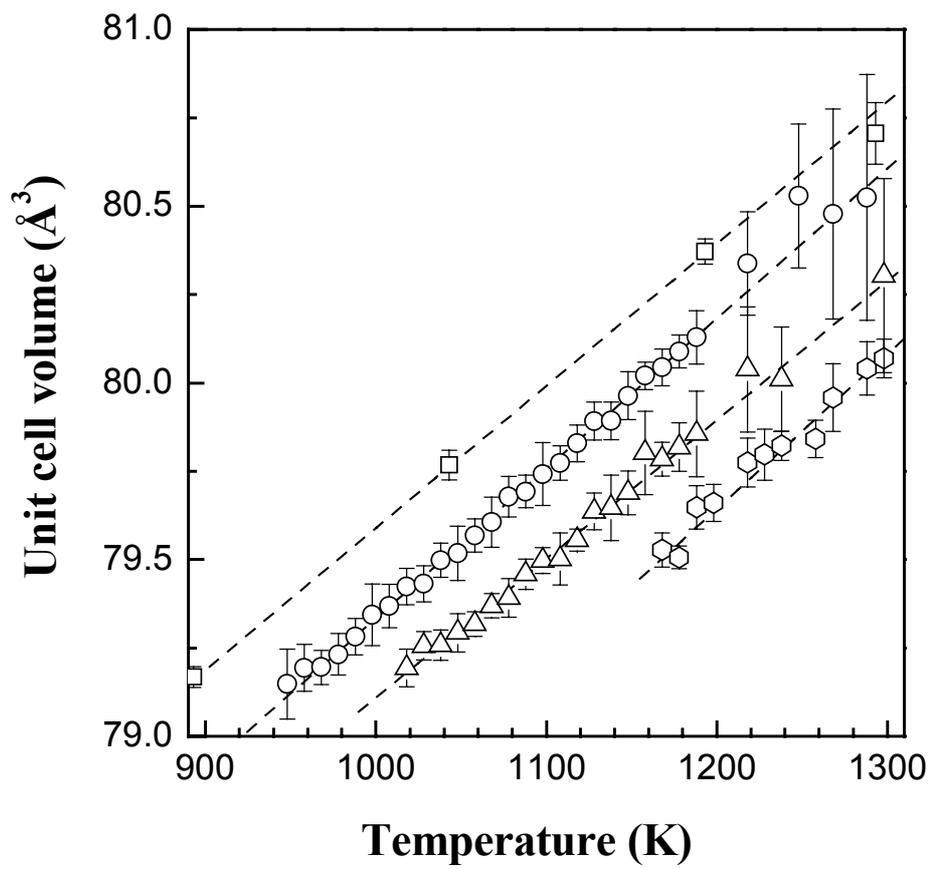

**Fig. 2**